\documentclass[epj,referee]{svjour}
\usepackage[english]{babel}
\usepackage{graphicx}
\usepackage{amsfonts}
\usepackage{amsmath}
\usepackage{amssymb}

\def\Journal#1#2#3#4#5{{#1}, #2 \textbf{#3}, #4 (#5)}

\newcommand{\SI}{S_0}
\newcommand{\xs}{k}
\newcommand{\Ps}{P_{\infty}}
\newcommand{\EE}{\mathbb{E}}

\newcommand{\Qm}{\mathbb{Q}}
\newcommand{\Pm}{\mathbb{P}}
\newcommand{\Po}{\Phi}
\newcommand{\uu}{\mathbf 1}
\newcommand{\CC}{\mathcal{C}\/}
\newcommand{\PP}{\mathcal{P}\/}
\newcommand{\FF}{\mathcal{F}\/}
\newcommand{\NN}{\mathcal{N}\/}
\newcommand{\MM}{\mathcal{M}\/}
\newcommand{\LL}{\mathcal{L}\/}
\newcommand{\dd}{\mbox{d}}

\begin{document}

\title{Renewal equations for option pricing}
\author{Miquel Montero}
\institute{Departament de F\'{\i}sica Fonamental, Universitat de Barcelona, Diagonal 647, E-08028 Barcelona, Spain.\\
\email{miquel.montero@ub.edu}}
\date{\today}
\abstract{
In this paper we will develop a methodology for obtaining pricing expressions for financial instruments whose underlying asset can be 
described through a simple continuous-time random walk (CTRW) market model. Our approach is very natural to the issue because it is based in the use of renewal equations, and therefore it enhances the potential use of CTRW techniques in finance. We solve these equations for typical contract specifications, in a particular but exemplifying case. We also show how a formal general solution can be found for more exotic derivatives, and we compare prices for alternative models of the underlying. Finally, we recover the celebrated results for the Wiener process under certain limits. 
}     
\maketitle

\section{Introduction and motivation}
The continuous-time random walk (CTRW) formalism, introduced in the physics literature by Montroll and Weiss~\cite{MW65}, is a way to generalize ordinary random walks by letting the steps and the time elapsed between them be random magnitudes. In this sense, CTRWs are related to several other well-known extensions of random walks in continuous time, like semi-Markov processes or Markov renewal processes~\cite{CM65}, the oldest of which is perhaps the pure-birth Poisson process~\cite{GUY24,WF50}. Physicists have extensively used CTRWs in the past in the study of a large variety of physical phenomena~\cite{W94}, and lately also in the modelling of financial data |see for instance~\cite{MMW03,MMPW06} or the exhaustive review by Scalas~\cite{ES06}.


A very distinctive feature of the CTRW formalism is that most of the statistical properties of these processes can be expressed in the form of renewal equations |the reader can find in ref.~\cite{C62} a classical introduction to renewal theory.
In this paper we will develop what it is perhaps the most natural approach for obtaining pricing expressions for financial instruments~\cite{W98} whose underlying can be described through a single CTRW. It is our belief that this methodology, based in the use of renewal equations, puts the issue under a different perspective, eases its interpretation in physical terms, and therefore enhances the potential extension of existing CTRW techniques and results to finance. 

The mere idea of considering a pure jump process as a market model is far from being new. It is due to Cox and Ross~\cite{CR76}, but it gained relevance after Merton's introduction of jump-diffusion models~\cite{M76}. Since then a lot of work has been made on the issue of financial processes with random jumps |see~\cite{CT04} for a comprehensive text on this matter. The usual interpretation of these models identifies such sudden changes with {\it abnormal\/} market behaviour, {\it e.g.\/} a crash, whereas diffusion determines the {\it normal\/} evolution: this perspective appears in the seminal paper by Merton~\cite{M76}, but it is still in use nowadays~\cite{AS04,ASSW06}. This is one of the motivations behind the use of L\'evy processes in the description of random jumps, because they provide a mechanism to incorporate the non-Gaussian behaviour observed in financial data into option pricing~\cite{BL02}. 

It is self-evident that one can obtain the pure jump counterpart of any result based on a market model that mixes a diffusive process with jumps, just by setting the diffusion coefficient equal to zero. However, we are interested in a description of the process where  jumps condense all the stochastic behaviour of the market, as in~\cite{R07}, what is not so usual in the literature. In fact, in some sense, our model is able to follow the opposite path: as we will show below, we can {\it recover\/} the Merton-Black-Scholes results for the Wiener process under certain limits.  


Another interesting property of the model is that, even though we will eventually restrict our study to the case in which waiting times are exponentially distributed |what turns our CTRW into an instance of compound Poisson process| the formalism admits of different sojourn distributions. 
Our aim at this point is to obtain sound financial results and, as we will discuss later, general CTRW market models
might allow for arbitrage possibilities: it would be feasible to obtain riskless earnings without net investment of capital. The absence of arbitrage opportunities is a cornerstone in option pricing theory but, in practical situations, a small level of arbitrage would be admissible provided that it does not overcome transactional costs, an interesting possibility to explore in the future. 

The paper is structured as follows. In section~\ref{sect_dev} we show how, in fact, the natural way for pricing derivatives when the stock price follows a CTRW process is through renewal equations. Basic ideas about the financial rigour of all the expressions are given, but we also devote appendix~A to discussing their ultimate mathematical foundations. Section~\ref{sect_eur} deals with European options. Here we present, on the one hand, exact pricing expressions for the most typical contract specifications, but for a particular choice of the distribution of jump sizes, and on the other hand, a formal solution valid for a general distribution, but restricted to more exotic derivatives. Section~\ref{sect_am} is devoted to American options, an elusive topic because for these derivatives few exact solutions are known. We revisit some of these instances, among which {\it perpetual\/} options deserve special attention. 
Finally, conclusions are drawn in section~\ref{sect_end}.      

\section{Derivatives and CTRW processes}
\label{sect_dev}
In the most common version of the CTRW, any realization of the process $X(t)$ consists of a series of step functions: it changes at random times $t_0, t_1, t_2,\cdots$ while it remains fixed in place between successive steps. The interval between these successive steps is the random variable $\Delta t_n=t_n-t_{n-1}$  called sojourn or waiting time. At the conclusion of the $n$th sojourn $X(t)$ experiences a random change, or jump, given by 
\begin{equation*}
\Delta X_n\equiv\Delta X(t_n)=X(t_n)-X(t_{n-1}).
\end{equation*}
Both waiting times $\Delta t_n$ and random jumps $\Delta X_n$ are assumed to be (mutually) independent and identically distributed random variables described by their probability density functions (PDFs) which we denote by $\psi(\cdot)$ and $h(\cdot)$ respectively. 

One of the main objectives within the CTRW framework is to obtain the so-called propagator, the transition PDF of $X(t)$, defined by
\begin{equation*}
p(x-x_0,t-t_0)\dd x=\Pr \{x<X(t)\leq x+\dd x|X(t_0)=x_0\}.
\end{equation*}
If $t_0$ is a jump time, as we will assume hereafter, the propagator obeys the following renewal equation~\cite{W94,MMW03}:
\begin{eqnarray*}
p(y,\tau)=\delta(y)\int^{\infty}_{\tau} \psi(\tau') \dd \tau'\\
+\int_{0}^{\tau} \dd \tau' \psi(\tau')\int_{-\infty}^\infty h(y')p(y-y',\tau-\tau')\dd y',
\end{eqnarray*}
|here $y= x-x_0$, and $\tau=t-t_0$. This integral equation can be solved in the Fourier-Laplace space:
\begin{equation}
\hat{\tilde p}(\omega,s) = \frac{1}{s}\frac{{1 - \hat\psi (s)}}{{1 - \tilde h(\omega)\hat\psi (s)}},
\label{hat_tilde}
\end{equation} 
where $\hat{\tilde p}(\omega,s)$ is the joint Fourier-Laplace transform of function $p(y,\tau)$, $\tilde h(\omega)$ is the Fourier transform of $h(y)$, and $\hat\psi (s)$ is the Laplace transform of $\psi(\tau)$. Similar notation is used along the text.

Let us assume that our random process is $X(t)=\ln S(t)$, where $S(t)>0$ is the price of some stock at time $t$, and concentrate our attention in the study of its derivatives |financial instruments whose value depends on (derives from) present and past states of the asset $S$, which is commonly referred as {\it the underlying\/}.
A typical example are European options: contracts between two parties that give the buyer the right, but not the obligation, to buy (call) or sell (put) shares of the underlying stock at some prearranged price, the {\it strike price\/}, on a specific date in the future, the maturity or expiration time $T$. The problem is in essence how to relate the present ($t_0$) value of the option $\CC(x_0,t_0)$, $x_0=\ln S_0=\ln S(t_0)$, which is unknown, with 
$\Po(x)$, the value of the option at expiration, which is fixed beforehand.

Note that, in absence of inflation, $\CC(x_0,t_0)$ should coincide with the {\it pay-off function\/} $\Po(x_0)$ if there was no event in the $T-t_0$ interval, whereas if the first jump moved the process from $x_0$ to $x_1$ at some instant $t_1$, $t_0 \leqslant t_1 \leqslant T$, the new price ought to be simply $\CC(x_1,t_1)$. Therefore, we can evaluate the likelihood of these disjoint scenarios, and put them together in order to obtain the renewal equation that the price must obey right after a jump:
\begin{eqnarray}
\CC(x_0,t_0) &=& \Po(x_0)e^{-r(T-t_0)} \int^{\infty}_{T} \dd t_1 \psi(t_1-t_0)\nonumber \\
&+& \int^{T}_{t_0}\dd t_1 \psi(t_1-t_0) e^{-r(t_1-t_0)}\nonumber \\ &\times& \int^{+\infty}_{-\infty}h(x_1-x_0) \CC(x_1,t_1)\dd x_1,
\label{ERenewal}
\end{eqnarray}
where the factors containing $r\geqslant 0$, the risk-free interest rate, take into account the natural depreciation of financial items as time passes. We must also recall that contract specifications do not only define the shape of the pay-off function but also provide complementary boundary conditions that the option price $\CC(x_0,t_0)$ has to fulfill. This additional constraints will completely determine the solution of~(\ref{ERenewal}). 
$\CC(x_0,t_0)$ will  also allow us to compute by integration the option price for any given time between jumps, $\CC(x_0,t_0+\tau;x_0,t_0)$, $0<\tau\leqslant T-t_0$, if we know when the last event took place and the actual stock price:
\begin{eqnarray*}
\CC(x_0,t_0+\tau;x_0,t_0) &=& \Po(x_0)e^{-r(T-t_0-\tau)} \frac{1-\Psi(T-t_0)}{1-\Psi(\tau)}\nonumber \\
&+& \int^{T}_{t_0+\tau}\dd t_1 \frac{\psi(t_1-t_0)}{1-\Psi(\tau)} e^{-r(t_1-t_0-\tau)} \nonumber \\ 
&\times& \int^{+\infty}_{-\infty}h(x_1-x_0) \CC(x_1,t_1)\dd x_1,
\end{eqnarray*}
where $\Psi(t)\equiv\int^{t}_{0} \dd t' \psi(t')$ is the cumulative distribution function of sojourn times.


Up to this point we have taken into consideration arguments based on the renewal properties of the process alone. This is not guaranteeing the validity from a financial point of view of the previous reasoning in general, and of eq.~(\ref{ERenewal}) in particular. The main object of concern is the issue of the efficiency of the market: inefficient markets allow for arbitrage opportunities, the possibility of obtaining riskless profits, which is not a desired feature of the model. We will present here a succinct description of consequences and constraints of the efficient market hypothesis, but the reader can find in appendix~A a detailed development of the mathematical theory behind. A sufficient condition to have an efficient market is that the discounted process $S(t) e^{-r t}$ fulfills the martingale property, which states  
\begin{equation}
\EE^{\Pm}[S(t') e^{-r t'}|\FF(t)]=S(t) e^{-r t},
\label{2aa}
\end{equation}
for any $t\leqslant t'$. Here $\FF(t)$ denotes all the available information up to time $t$ and the superscript $\Pm$ indicates that we are using the {\it physical\/} measure, the one that describes the probabilistic properties of the actual process $S(t)$.~\footnote{$\FF(t)$ corresponds to what is known in mathematical terminology as a filtration: an increasing one-parameter family of sub $\sigma$-algebras of $\FF$, $\FF(t)\subseteq \FF(t') \subseteq \FF$, $t \leqslant t'$, where $\FF$ is a $\sigma$-algebra of subsets of the sample space $\Omega$. The sample space, $\FF$ and the measure $\Pm$ define the probability space ($\Omega,\FF,\Pm$).} 

Let us show now that when $r\neq 0$ and $\EE^{\Pm}[e^{\Delta X_n}]=\tilde{h}(\omega=-i)\neq 1$,  waiting times must be exponentially distributed. Assume that eq.~(\ref{2aa}) is true for any $t$ and $t'$, and in particular that it holds for $t=t_0$, a jump time:
\begin{eqnarray*}
\SI e^{-r t_0}&=&e^{x_0}e^{-r t_0}=\EE^{\Pm}[S(t') e^{-r t'}|\FF(t_0)]\\&=&\int^{+\infty}_{-\infty} e^x  e^{-r t'} p(x-x_0,t'-t_0)\dd x,
\end{eqnarray*}
therefore
\begin{eqnarray*}
  e^{r (t'-t_0)}=\int^{+\infty}_{-\infty} e^{x-x_0} p(x-x_0,t'-t_0)\dd x.
\end{eqnarray*}
Let us multiply this identity by $e^{-s(t'-t_0)}$, $s>r$, and integrate it for any value of $t'$, $t'\geqslant t_0$:
\begin{eqnarray*}
\frac{1}{s-r}&=&\int^{+\infty}_{-\infty} e^{x-x_0} \hat p(x-x_0,s)\dd x=\hat{\tilde p}\left(\omega=-i,s\right) \\&=& \frac{1}{s}\frac{{1 - \hat\psi (s)}}{{1 - \tilde h(-i)\hat\psi (s)}},
\end{eqnarray*}
where we have used~(\ref{hat_tilde}). Therefore
\begin{eqnarray*}
\hat\psi (s)=\frac{r}{s \left[\tilde h(-i)-1\right]+r}\quad (s>r).
\end{eqnarray*}
Since $\psi(\tau)$ is a PDF its Laplace transform $\hat\psi (s)$ is a smooth funcion on $s\geqslant 0$, and the previous expression must be true for any value of $s$. Then $\psi(\tau)=\lambda e^{-\lambda \tau}$, with intensity $\lambda$,
\begin{equation}
\lambda=\frac{r}{\tilde{h}(-i)-1}.
\label{lambda}
\end{equation}
Note that if $r>0$, which is the most usual situation, $1<\tilde{h}(-i)<\infty$ in order to keep $0<\lambda<\infty$. In spite of the fact that we have assumed that $t=t_0$ is a jump time, the martingale property holds for any $t$ in this case, since when sojourns are exponentially distributed, the jump occurrence follows a Poisson process, and therefore any time instant $t$ can be thought as a renewal point of the prosses $X(t)$. Let us finally mention that when the jump size density is such that 
$\tilde{h}(-i)=1$ and $r=0$, {\it i.e.\/} if the economy is strictly neither inflationary nor deflationary, 
the martingale condition is identically satisfied and the sojourn distribution becomes arbitrary~\cite{W91}. Since we are not planning to study this case here, we will consider hereafter that our process is a compound Poisson process.

Condition~(\ref{2aa}) is a sufficient condition but not a necessary condition to have an efficient market model. In fact, the necessary and sufficient condition is that one can define a risk-neutral market measure $\Qm$, different from $\Pm$ but describing the same kind of process, for which the martingale condition holds. In our case, from a practical point of view, this means that we have to replace the actual intensity of the Poisson process with the risk-neutral intensity~(\ref{lambda}).

In order to follow the development up, it is very convenient to work with the backward version of eq.~(\ref{ERenewal}) by introducing $C(x,\bar{t})=\CC(x,T-\bar{t})$ and $\bar{t}=T-t$:
\begin{eqnarray}
C(x,\bar{t}) = \Po(x)e^{-(\lambda+r) \bar{t}} \nonumber \\
+ \lambda \int^{\bar{t}}_{0} \dd \bar{t}' e^{-(\lambda+r)(\bar{t}-\bar{t}')}\int^{+\infty}_{-\infty} h(y-x) C(y,\bar{t}')\dd y.
\label{ERenewalB}
\end{eqnarray}
One can show that 
$C(x,\bar{t})$ fulfills the classical Merton's equation~\cite{M76} for jump-diffusive market models once one removes the contribution of the Wiener process to the asset evolution, as we have pointed before: the perspective is different but any result must be consistent with previous developments.

\section{European option prices}
\label{sect_eur}
\subsection{Exact solutions for a particular case}
We have several alternative ways to follow in order to solve eq.~(\ref{ERenewalB}). The traditional method would transform~(\ref{ERenewalB}) into a partial integro-differential equation~\cite{CT04}. Another approach is to move it into the Fourier-Laplace domain and thus obtain a formal solution to the problem that will be valid for a general $h(x)$~\cite{BL02}. This procedure is delicate because in general neither $\tilde{\Po}(\omega)$ nor $\hat{\tilde{C}}(\omega,s)$ 
do properly exist. However, one can avoid this problem by assuming that $\omega$ is a complex variable. We will show the outcomes of this methodology in the next section in a case in which $\tilde{\Po}(\omega)$ is regular. There we will consider a jump distribution that connects our development with the issue of L\'evy flights, and therefore with L\'evy processes in general. In such a situation the nature of the L\'evy density makes feasible to write~(\ref{ERenewalB}) as a fractional partial differential equation~\cite{SGM00,CC07}. 
Within our approach we will avoid turning eq.~(\ref{ERenewalB}) into a partial differential equation that involves time derivatives. This choice 
eludes the ambiguity that may appear in the meaning of time derivatives |time evolution versus parametric dependence| 
when one has different pricing expressions, depending on whether present time coincides with a jump or not.

To this end,
in the example we are presenting in this section, 
the asymmetric two-sided exponential PDF $h(x)$,~\footnote{Along the text $\uu_{\{\cdot\}}$ will denote the indicator function, which assigns the value 1 to a true statement, and the value 0 to a false statement.} 
\begin{equation}
h(x)=\frac{\gamma \rho}{\gamma + \rho} \left[e^{-\rho x} \uu_{x\geqslant 0} + e^{\gamma x} \uu_{x<0}\right] \quad \left(\rho>0,\gamma>0\right),
\label{h}
\end{equation}
we will derive a second-order {\it ordinary\/} differential equation for Laplace transform of the option price:
\begin{eqnarray}
\hat{C}(x,s) =\nonumber \\ 
\frac{1}{\lambda+r+s}\left\{ \Po(x) + \lambda \int^{+\infty}_{-\infty}\dd y h(y-x) \hat{C}(y,s)\right\}.
\label{ERenewalBL}
\end{eqnarray}
To the best of our knowledge this case, which possesses interesting properties as we shortly show, has not been previously analysed in the literature. Moreover, the choice is not arbitrary from the point of view of econometrics. Notwithstanding the extensive literature reporting that the occurrence of large changes in many financial data series presents a power-law decay, {\it e.g.\/} \cite{GPAMS99}, the empirical analysis of the distribution of single trade returns is a topic relatively unexplored~\cite{RE05}. An indirect evidence supporting our assumption on the shape of $h(x)$ comes from the increasing number of recent works concluding that at small time scales, moderate returns are better described through an exponential PDF |see~\cite{SY07} and references therein. 
Anyway, we will consider alternative functional forms for $h(x)$ in the next section.

After the choice in~(\ref{h}), the risk-free value of $\lambda$ reads
\begin{equation*}
\lambda=r\frac{(\rho-1)(\gamma+1)}{\gamma-\rho+1},
\end{equation*}
and the constraint $1<\tilde{h}(-i)<\infty$ implies $0<\rho-1<\gamma$. As we have announced above, in this case the integral eq.~(\ref{ERenewalBL}) transforms {\it de facto\/} into a second-order ordinary differential equation for $\hat{C}(x,s)$:
\begin{eqnarray*}
\partial^2_{xx}\hat{C}(x,s)+(\gamma-\rho)\partial_{x}\hat{C}(x,s)-\frac{r+s}{\lambda+r+s}\gamma\rho\hat{C}(x,s)=\\
\frac{1}{\lambda+r+s}\left\{\partial^2_{xx}\Po(x)+(\gamma-\rho)\partial_{x}\Po(x)-\gamma \rho \Po(x)\right\}. 
\end{eqnarray*}
The general solution of this differential equation is 
\begin{eqnarray}
\hat{C}(x,s)=A_+(s)e^{\beta_+ x}+A_-(s)e^{\beta_- x}+\frac{1}{\lambda+r+s}\Po(x) \nonumber \\
-\frac{\lambda\gamma\rho}{(\lambda+r+s)^2}\int^{x}\dd y \Po(y)\left[\frac{e^{\beta_+(x-y)}-e^{\beta_-(x-y)}}{\beta_+-\beta_-}\right], 
\label{general}
\\
\beta_{\pm}=-\frac{\gamma-\rho}{2} \pm \frac{1}{2}\sqrt{(\gamma+\rho)^2-\frac{4 \lambda \gamma \rho}{\lambda+r+s}} \gtrless 0, \nonumber
\end{eqnarray}
provided that $\Po(x)$ is a smooth-enough function. In practice, pay-off functions show at least one point where the second derivative is not well defined, let say $\xs=\ln K$. Then we have different solutions for the different regions. Consider for instance call options, where $\Po(x)=0$ for $x<\xs$. Since the option price must fulfill in this case that $\lim_{x\rightarrow -\infty} C(x,\bar{t})=0$, we will have,
\begin{eqnarray*}
\hat{C}(x,s)=A_+(s)e^{\beta_+ x}\quad (x<\xs). 
\end{eqnarray*}
The value of the pay-off function for $x \geqslant \xs$ is different for different option flavours. Binary call properties follow from the fact that $\Po(x)=1$ and the boundary condition $\lim_{x\rightarrow +\infty} C(x,\bar{t})=e^{-r\bar{t}}$. Then 
\begin{eqnarray*}
\hat{C}(x,s)=A_-(s)e^{\beta_- x}+\frac{1}{r+s}\quad (x \geqslant \xs). 
\end{eqnarray*}
Note that, like the process itself, option prices are discontinuous in general, and therefore we must use eq.~(\ref{ERenewalBL}) in order to determine functions $A_{\pm}(s)$:
\begin{eqnarray*}
A_{\pm}(s)=-\frac{\beta_{\mp}}{\beta_+-\beta_-}\frac{\lambda}{(\lambda+r+s)(r+s)} e^{-\beta_{\pm} \xs}.
\end{eqnarray*}
Now we can perform the Laplace inversion to get:
\begin{eqnarray*}
C(x,\bar{t})&=&\Bigg[\uu_{x\geqslant \xs}+ 2\int^{\infty}_0 \dd u I_1(2 u) \exp\left({-\frac{u^2}{\lambda \bar{t}}}\right) \\ &\times&  
\NN\left(\frac{x-\xs}{2 u} \sqrt{2 \gamma \rho \lambda \bar{t}}+\frac{\gamma-\rho}{\sqrt{2 \gamma \rho \lambda \bar{t}}} u\right)\Bigg]e^{-(\lambda+r)\bar{t}},
\end{eqnarray*}
where $\NN(\cdot)$ is the cumulative distribution function of a zero-mean unit-variance Gaussian PDF, and $I_n(\cdot)$ is a $n$-order modified Bessel function of the first kind. 

\begin{figure}
{\hfil 
\includegraphics[width=1.0\columnwidth,keepaspectratio=true]{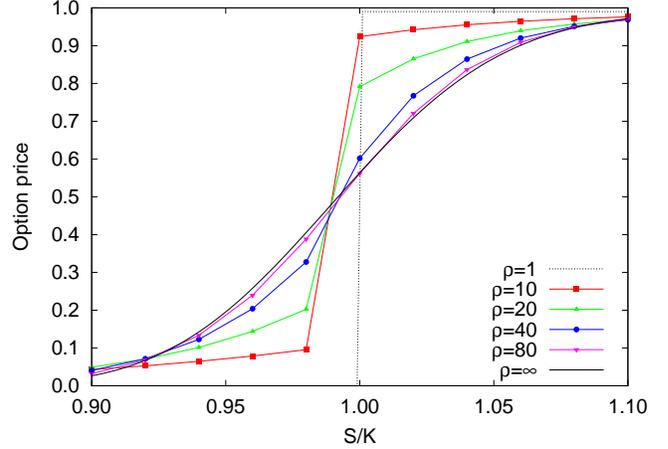}}
\caption{Price for an European binary call option. We have set $\gamma=\rho-1+2r/\sigma^2$, and we have chosen typical market values for $r=4\%$, $\sigma=10\%$, and $T-t_0=0.25$ years.}
\label{Fig1}
\end{figure}

Pay attention to the $\uu_{x\geqslant \xs}$ term. 
It counts for the finite possibility that the system keeps in place until expiration. In fig.~\ref{Fig1} we can see the how the relative contribution of this term diminishes for larger values of $\lambda$.  Indeed, the discontinuity disappears when considering continuous trading, $\lambda\rightarrow \infty$. We can approach to this limit by letting $\rho \rightarrow \infty$ and $\gamma \rightarrow \infty$, but in such a way that the difference remains finite $\infty>\gamma - \rho+1\equiv\varepsilon>0$. In fact we can relate $\sigma=\sqrt{2r/\varepsilon}$ with the volatility of the market, since
\begin{eqnarray*}
m_1(t-t_0)&\equiv&
\EE[X(t)-x_0|\FF(t_0)]
\\&=& \frac{\gamma-\rho}{\gamma \rho}\lambda (t-t_0)\rightarrow\left(r-\frac{1}{2}\sigma^2\right) (t-t_0),\\
m_2(t-t_0)&\equiv&\EE[(X(t)-x_0)^2|\FF(t_0)]-[m_1(t-t_0)]^2\\
&=&2\frac{\gamma^2-\gamma \rho+\rho^2}{\gamma^2 \rho^2}\lambda (t-t_0) \rightarrow \sigma^2 (t-t_0), 
\\
C(x,\bar{t})&\rightarrow &e^{-r\bar{t}} \NN\left[\frac{x-\xs+(r-\sigma^2/2)\bar{t}}{\sigma \sqrt{\bar{t}}}\right],
\end{eqnarray*}
the well-known results for a Wiener process~\cite{W98}.

\begin{figure}
{\hfil 
\includegraphics[width=1.0\columnwidth,keepaspectratio=true]{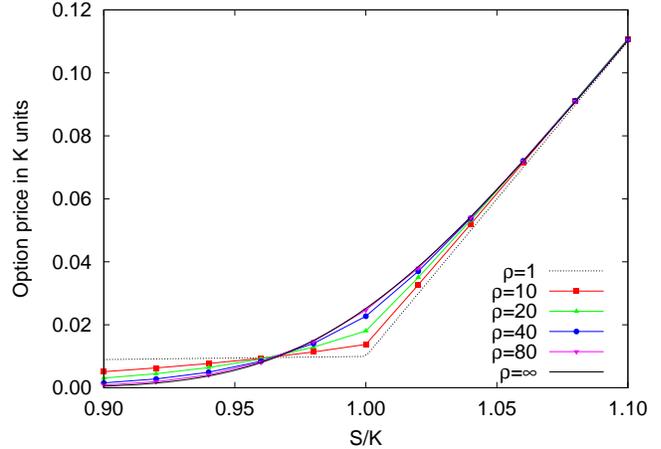}}
\caption{Price for an European vanilla call option. We have kept the same parameters as in fig.~\ref{Fig1}. Observe how option prices fit between the two limiting curves.}
\label{Fig2}
\end{figure}

Consider now the case of a vanilla call for which we have $\Po(x)=e^x-K$ when $x \geqslant \xs$, 
and
\begin{equation*}
\lim_{x\rightarrow +\infty} \frac{C(x,\bar{t})}{e^x- K e^{-r\bar{t}}}=1. 
\end{equation*}
the solution of the differential equation is:
\begin{eqnarray*}
\hat{C}(x,s)=A_-(s)e^{\beta_- x}+\frac{e^x}{s}-\frac{K}{r+s}\quad (x\geqslant \xs). 
\end{eqnarray*}
and functions $A_{\pm}(s)$ are in this case:
\begin{eqnarray*}
A_{\pm}(s)=\frac{1}{\lambda+ r+s}\left[ \frac{\lambda}{r+s} \beta_{\mp} + \frac{\lambda+r}{s}(1-\beta_{\mp})\right]\frac{e^{(1-\beta_{\pm}) \xs}}{\beta_+-\beta_-}.
\end{eqnarray*}
The Laplace inversion of $\hat{C}(x,s)$ is cumbersome
\begin{eqnarray*}
C(x,\bar{t})= \Bigg\{ 2\int^{\infty}_0 \dd u I_1\left(2u\right)
\Bigg[e^x \MM^{\gamma+1}_{\rho-1}\left(x-\xs;\sqrt{\frac{2u^2}{\gamma \rho \lambda\bar{t}}} \right)
\\- K \MM^{\gamma}_{\rho}\left(x-\xs;\sqrt{\frac{2u^2}{\gamma \rho \lambda\bar{t}}}\right)\Bigg]+\left[e^x-K\right]\uu_{x\geqslant \xs}\Bigg\}e^{-(\lambda+r)\bar{t}},\\
\MM^{a}_{b}\left(c;\xi\right)=e^{-a b \frac{\xi^2}{2}} \NN\left(\frac{a-b}{2} \xi+\frac{c}{\xi}\right),
\end{eqnarray*}
but still readable. In particular one can foresee how the classical Black-Scholes (BS) solution~\cite{BS73} appears in the continuous trading limit, see fig.~\ref{Fig2}. 
Also in this figure we observe that in this case the no-trade limit, which corresponds to $\rho \rightarrow 1$, 
leads to a non-trivial result:
\begin{eqnarray*}
C(x,\bar{t})&=&e^x\left(1-e^{-r \bar{t}}\right) + \left(e^x-K\right)e^{-r \bar{t}}\uu_{x\geqslant \xs}.
\end{eqnarray*}
This expression can aid in the qualitative analysis of the pricing curves. The only crossing of the no-trade solution ($\rho \rightarrow 1$) with the continuous trading solution ($\rho \rightarrow \infty$) marks the price-strike ratio (the {\it moneyness\/}) for which the result is less sensitive to the actual value of $\rho$, and so any solution apparently traverses this crossing |see again fig.~\ref{Fig2}.
When the moneyness is well below than one (a out-of-the-money option) the BS solution underestimate the CTRW price for any $\rho$ value. It is also noticeable that when the moneyness is about one (at-the-money options) the BS solution overestimate the CTRW, but it is not clear in fig.~\ref{Fig2} if the relative behaviour reverses one more time.

\begin{figure}
{\hfil 
\includegraphics[width=1.0\columnwidth,keepaspectratio=true]{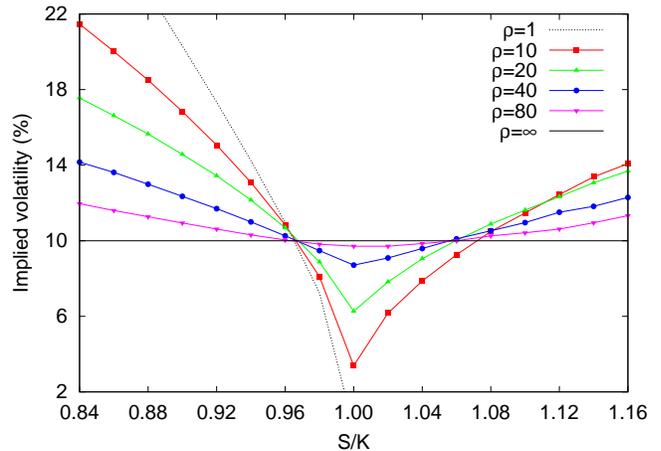}}
\caption{Implied volatilities for an European vanilla call option. The models coincide with those in fig.~\ref{Fig2}. The implied volatility of all solutions collapses for  $S \approx 0.965 K$. The curves turn upward for $S$ values larger than $K$, showing what is known as {\it volatility smile\/}, but they do not cross all together again.}
\label{IV1}
\end{figure}

The picture is more clear when depicted not in terms of prices but in terms of implied volatilities: the value for the volatility that one must introduce in the BS solution in order to reproduce a given option price. When all the parameters are kept unchanged, the BS price is a monotone increasing function of the volatility. In fig.~\ref{IV1} we observe how the implied volatility moves upward for options with a higher-than-one moneyness (in-the-money options) but, in contrast to the out-of-the money case, the crossing point with the BS value depends stronger on $\rho$.

\begin{figure}
{\hfil 
\includegraphics[width=1.0\columnwidth,keepaspectratio=true]{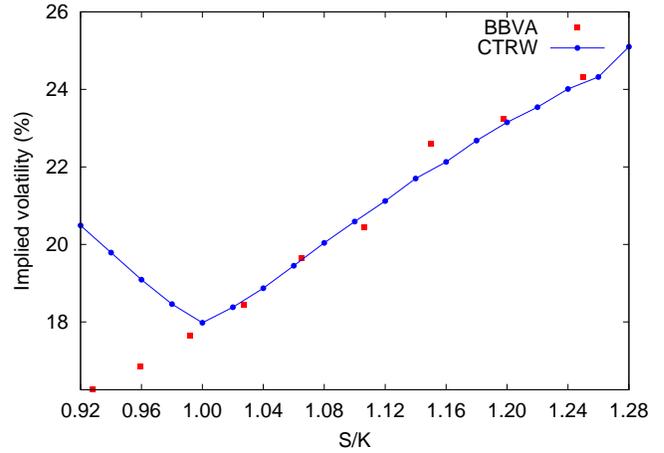}}
\caption{Comparison between implied volatilities for an European vanilla call option. The boxes show the implied volatilities of several American-style contracts whose underlying is the Spanish bank BBVA. The analysed day was October 17, 2005, 60 days before maturity: December 16, 2005. The spot price was 14.38 euros at the close of the market. The risk-free interest rate was assumed to be equal to the two-month Euribor on the selected date $r=2.139\%$. No dividend was paid during the observational period, what equalizes American and European call prices. We show how a CTRW model with $\rho=30$ and $\sigma=20\%$ may reproduce the upward slope of in-the-money calls. }
\label{IV2}
\end{figure}

This pattern in the implied volatility plot, in which at-the-money options tend to have lower implied volatilities than other options, is known as {\it the volatility smile\/}, a well-known empirical phenomenon that gained relevance after the October 1987 crash. This stilized  fact is ubiquitous in any financial derivative but it appears with different flavours depending on the nature of the underlying asset. In particular, equity options tend to show nowadays an upward sloping implied volatility curve,~\footnote{When one analyses empirical data it is somewhat more natural to plot implied volatility as a function of $K/S$ rather than $S/K$, because $S$ is fixed at closing time and one has different option prices for different strike values. Obviously, in terms of $K/S$ the curve is downward sloping.} {\it i.e.\/} a volatility {\it skew\/}, as we can see in fig.~\ref{IV2}. Superimposed with this practical example we find the implied volatility curve corresponding to a CTRW market model with the round values of $\rho=30$ and $\sigma=20\%$. Therefore this model is amenable enough to (partially) reproduce actual data behaviour.

Finally, note that the previous results for European calls can be used in order to obtain put prices, $P(x,\bar{t})=\PP(x,T-\bar{t})$, because the so-called put-call parity stands also in our case:
\begin{eqnarray*}
P(x,\bar{t}) + C(x,\bar{t})=e^{-r\bar{t}}& & \mbox{ (binary),} \\ 
P(x,\bar{t}) - C(x,\bar{t})=K e^{-r\bar{t}}-e^x& & \mbox{ (vanilla).} 
\end{eqnarray*}
This statement can be proven by using eq.~(\ref{general}) for $\hat{F}(x,s)=\hat{P}(x,s)+\hat{C}(x,s)$ (binary) and $\hat{F}(x,s)=\hat{P}(x,s)-\hat{C}(x,s)$ (vanilla) because $\Po(x)$ is regular: $\Po(x)=1$ and $\Po(x)=K-e^x$, respectively.

\subsection{General solution for integrable pay-offs}
\begin{figure}
{\hfil 
\includegraphics[width=1.0\columnwidth,keepaspectratio=true]{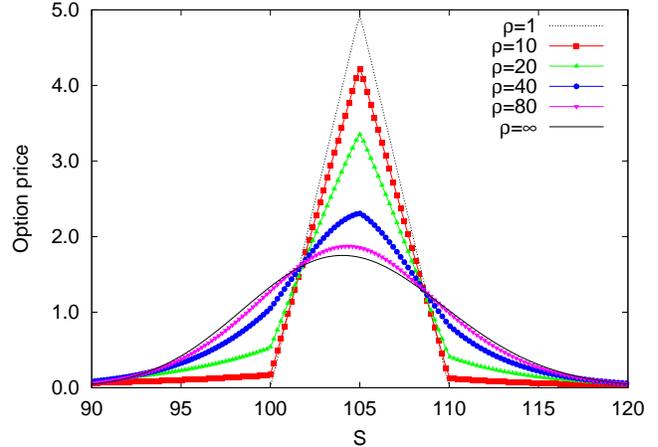}}
\caption{Price for the European portfolio~(\ref{portfolio}). We have set $K=100$, $L=10$, and kept the rest of parameters as in fig.~\ref{Fig1}. The value of $\rho$ changes drastically the value of the position.}
\label{Fig3}
\end{figure}

When $\tilde \Po(\omega)$ exists we can move eq.~(\ref{ERenewalBL}) into Fourier domain by defining the joint Fourier-Laplace transform of the option price, $\hat{\tilde C}(\omega,s)$. 
One can show that this magnitude fulfills:
\begin{eqnarray*}
\hat{\tilde{C}}(\omega,s) =\frac{ \tilde{\Po}(\omega)}{s+r+\lambda[1-\tilde{h}(-\omega)]},
\label{FL}
\end{eqnarray*}
and therefore the general solution of the problem follows:
\begin{eqnarray}
C(x,\bar{t}) 
&=&\frac{1}{2 \pi} \int^{+\infty}_{-\infty} \dd \omega {\tilde \Po}(\omega) {\tilde p}(-\omega,\bar{t}) e^{-i \omega x} \nonumber \\
&=&\frac{1}{2 \pi} \int^{+\infty}_{-\infty} \dd \omega {\tilde \Po}(\omega) e^{-\left\{r+\lambda[1-\tilde{h}(-\omega)]\right\}\bar{t}-i \omega x}.
\label{formal}
\end{eqnarray}
In this case, for a given pay-off function, we may analyse the dependence of the option price on the shape of $h(x)$ by means of numerical techniques (at least) as in~\cite{BL02}. Let us consider, for instance, the following pay-off: 
\begin{equation}
\Po(x)=(e^x-K)\uu_{k_1 \leqslant x \leqslant k_2} +(K+L-e^x)\uu_{k_2 < x \leqslant k_3},
\label{portfolio}
\end{equation} 
with $k_1=\ln(K)$, $k_2=\ln(K+L/2)$, and $k_3=\ln(K+L)$, $L>0$. This pay-off function may resemble bizarre but corresponds to the portfolio that results after buying two vanilla options with strike price $K+L/2$ and selling two more calls, one with strike $K$ and the other with strike $K+L$. In this sense, the value of the position when we know the appropriate expression for the vanilla call price is a simple question of arithmetic. In sum, we have a continuous~\footnote{This property is desirable from a practical point of view, because it avoids the annoying presence of the Gibbs phenomenon when one computes Fourier transforms numerically.} pay-off function whose Fourier transform reads:
\begin{equation*}
{\tilde \Po}(\omega)=\frac{1}{\omega(\omega -i)} \left[2 e^{(1+i\omega)k_2}- e^{(1+i\omega)k_1} -e^{(1+i\omega)k_3} \right], 
\end{equation*}
and for which we can compute the value of $C(x,\bar{t})$. In particular we can check the outcome that corresponds to the case we have extensively analysed in the previous section. In fig.~\ref{Fig3} we observe again noticeable divergences when assuming different values for $\rho$.

\begin{figure}
{\hfil
\includegraphics[width=1.0\columnwidth,keepaspectratio=true]{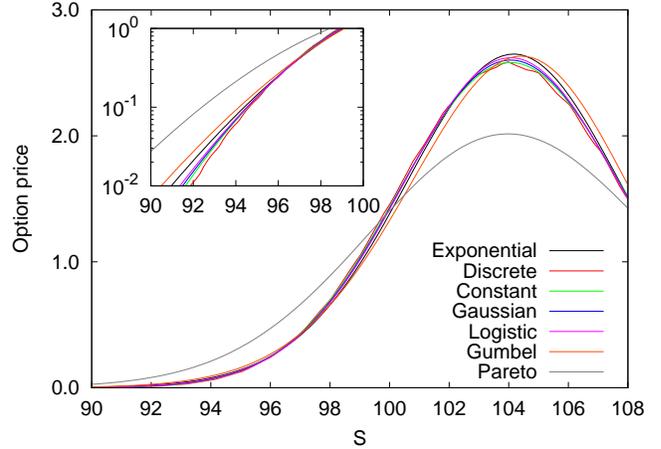}
}
\caption{Price of the European portfolio in~(\ref{portfolio}), for  the jump densities in tab.~\ref{Tab_dens}, with $\mu_1=10^{-3}$, $\mu_2=10^{-4}$, $r=4\%$ and $T-t_0=0.25$ years. Discrepancies induced by the effect of different skewness (and higher moment) values are noticeable.}
\label{Fig4}
\end{figure}

We must recall at this point that eq.~(\ref{lambda}) ensures the same (risk-neutral) market behaviour, in spite of the value of $\rho$. It is obvious that we have chosen the set of $\rho$ values in the previous figures on the basis of illustrative purposes. It is self-evident that it is more likely that actual market conditions lead to $\rho \gg 1$ than $\rho \sim 1$. Since for large values of $\rho$ the price converges to the Wiener result, one could argue against the practical relevance of the shape of $h(x)$. Eq.~(\ref{formal}) opens the possibility of exploring this issue systematically. Let us assume that we empirically determine the mean and the variance of density $h(x)$:
\begin{equation*}
\mu_1=\int^{+\infty}_{-\infty} x h(x) \dd x, \quad \mu_2=\int^{+\infty}_{-\infty} (x-\mu_1)^2 h(x) \dd x,
\end{equation*}
and consider the universe of two-parameter PDFs, like~(\ref{h}) which for comparative purposes we will rewrite in what remains of the section as
\begin{equation*}
h(x)=\frac{1}{a+b} \left[e^{-x/a} \uu_{x\geqslant 0} + e^{x/b} \uu_{x<0}\right].
\end{equation*}
It is obvious that $\mu_1$ and $\mu_2$ will completely settle $h(x)$. In table~\ref{Tab_dens} we present several candidates together with some relevant information. 
\begin{table*}
\begin{center}
\begin{tabular}{c c c c c}
\hline \hline
&$h(x)$&$\tilde{h}(\omega)$&$\mu_1$&$\mu_2$\\ \hline  
Exponential &
$\frac{1}{a+b} \left[e^{-x/a} \uu_{x\geqslant 0} + e^{x/b} \uu_{x<0}\right]$ &
$\frac{1}{(1-i\omega a)(1+i\omega b)}$ &
$a-b$ &
$a^2+b^2$ \\
Discrete &
$a \delta(x-b) + (1-a) \delta(x+b)$ &
$a e^{ib\omega}+(1-a) e^{-ib\omega}$ &
$(2a -1)b$ &
$4a(1-a)b^2$ \\
Constant &
$\frac{1}{b-a}\uu_{a\leqslant x \leqslant b}$ &
$\frac{i}{(b-a)\omega}\left( e^{ia\omega}- e^{ib\omega}\right)$ &
$\frac{b +a}{2}$ &
$\frac{(b-a)^2}{12}$ \\
Gaussian &
$\frac{1}{\sqrt{2 \pi b^2}} e^{-\frac{(x-a)^2}{2 b^2}}$ &
$e^{-\frac{b}{2}^2 \omega^2 + i a \omega}$ &
$a$ &
$b^2$ \\ 
Logistic &
$\frac{1}{4 b}\mbox{sech}^2\left(\frac{x-a}{2 b}\right)$ &
$e^{i a \omega}\Gamma(1-i b\omega) \Gamma(1+i b\omega)$&
$a$ &
$\frac{\pi^2}{3}b^2$ \\
Gumbel  &
$\frac{1}{b}e^{-\frac{x-a}{b}-e^{-(x-a)/b}}$ &
$e^{i a \omega}\Gamma(1-i b\omega)$ &
$a-b\Gamma'(1)$ &
$\frac{\pi^2}{6}b^2$ \\ 
Pareto  &
$\frac{\sqrt{b}}{|x|^{3/2}} [a + (1-2a)\uu_{x< 0}]e^{-\frac{|x|}{b}}$ &
See main text&
$\sqrt{\pi}(2a-1)b$ &
$\frac{\sqrt{\pi}}{2}b^2-\mu_1^2$ \\
\hline \hline
\end{tabular}
\end{center}
\caption{Definitions and properties of the different densities we consider in fig.~\ref{Fig4}. $\Gamma(\cdot)$ is the gamma function, and $\Gamma'(1)$ is minus the Euler-Mascheroni constant.}
\label{Tab_dens}
\end{table*}
In order to compare the different outcomes we have selected two round quantities, $\mu_1=10^{-3}$ and $\mu_2=10^{-4}$, that are far more plausible values for these magnitudes,~\footnote{For instance, for a Discrete $h(x)$, we will have $a\simeq 55\%$ and $b\simeq 1\%$. These numbers  may correspond to a stock that quotes about 100 times over the minimum tick change, in a bullish market.} and for which the Wiener limit is almost attained in general. In fig.~\ref{Fig4} we observe how intrinsic properties of the PDF functions, like the skewness or the excess of kurtosis, play a role that may cause the price to increase in a sensible amount: under the analysed conditions the Gumbel price triplicates that of Discrete when $S=92$. 

The sensibility of the result with respect to the shape of $h(x)$ is magnified if the jump PDF is such that $p(x,t)$ has not converged (enough) to a Gaussian. Consider for instance the case of a truncated Pareto, 
\begin{equation}
h(x)=\frac{b^{\nu}}{|x|^{1+\nu}} [a + (1-2a)\uu_{x< 0}]e^{-\frac{|x|}{b}},
\label{Pareto_h}
\end{equation}
which leads to the typical propagator of a truncated L\'evy flight process~\cite{MS94,K95},  also named as KoBoL process in the mathematical literature~\cite{BL02,CC07}. Similar results would have been obtained if we had used a CGMY process~\cite{CGMY02} 
\begin{equation*}
h(x)=\frac{a^{\nu}}{2 |x|^{1+\nu}}e^{-\frac{|x|}{a}}\uu_{x \geqslant 0} + \frac{b^{\nu}}{2|x|^{1+\nu}}e^{-\frac{|x|}{b}}\uu_{x< 0}.
\end{equation*}
The significance of this kind of distributions in the analysis of financial problems has been reported in the past |see for instance~\cite{BL00,BL02b} or~\cite{ML07} for a more recent work. However, this fact does not invalidate the rest of candidates because there are also evidences~\cite{MMW03,ES06} supporting the conclusion that in some markets a power-law decay in $p(x,t)$ may be the consequence of a scale-free behaviour in $\psi(t)$ rather that in $h(x)$. And the replacement of our Poisson process with a point process with a different waiting-time density is a much more delicate issue from the financial point of view, as we have argued above. Alternatively, a suggested way to follow is to extend the memory of the process by letting the intensity $\lambda$ be a function on past waiting time values: this keeps the martingale property of compensated sojourns $\Delta \bar{t}_n$, $\Delta \bar{t}_n=\Delta t_n-\EE[\Delta t_n|\FF(t_{n-1})]$, and it is compatible with different functional forms for the unconditional waiting-time PDF~\cite{ER98}. 

From definition~(\ref{Pareto_h}) we can compute the values of both $\mu_1=(2a-1) b \Gamma(1-\nu)$ and $\mu_2=b^2 \Gamma(2-\nu)-\mu_1^2$, whenever $0\leqslant \nu <1$, even though $h(x)$ is {\it not\/} a density since it is not integrable.~\footnote{This fact implies that, in spite of the way in which we have introduced them, from a mathematical point of view, KoBoL and CGMY processes with $\nu>0$ are pure jump L\'evy processes but not compound Poisson processes because the number of jumps in a finite time interval is infinite.} Note however that we need $1-\tilde{h}(-\omega)$, which formally equals:
\begin{equation*}
1-\tilde{h}(-\omega)=\int^{+\infty}_{-\infty} \left(1-e^{-i\omega x}\right)h(x)\dd x,
\end{equation*}
therefore we compute $\tilde{h}(\omega)$ as
\begin{equation*}
\tilde{h}(\omega)=1-\int^{+\infty}_{-\infty} \left(1-e^{i\omega x}\right)h(x)\dd x,
\end{equation*}
which is regular if the Pareto index $\nu$ is again in the range $0\leqslant \nu <1$.~\footnote{For $1\leqslant \nu <2$ some additional regularization must be done. We will not discuss it here.} Note that this procedure does not change the value of $\mu_1$ and $\mu_2$.
The general form of $\tilde{h}(\omega)$ for this case can be found in ref.~\cite{K95}, but, in particular, when the index is $\nu=1/2$, it reads:
\begin{equation*}
\tilde{h}(\omega)=1-2\sqrt{\pi}\left[a\sqrt{1-i \omega b}+(1-a)\sqrt{1+i \omega b}-1\right].
\end{equation*}
This means that $\lambda$ must be set as:
\begin{equation*}
\lambda=\frac{r}{2 \sqrt{\pi} \left[a(1-\sqrt{1-b})+(1-a)(1-\sqrt{1+b})\right]}.
\end{equation*}
The slow convergence of the truncated L\'evy flight PDF to a Gaussian promotes a very different price in this case, as can be observed in fig.~\ref{Fig4}.

\section{American put options}
\label{sect_am}
Let us present next the renewal formulae for American options, derivatives that can be exercised at any time before expiration. When dealing with American derivatives it is more convenient that we focus our attention on puts rather than calls because when the stock pays no dividends it can be shown that American calls are never early exercised and therefore become European options. The (backward) renewal equation that follow live American put options is
\begin{eqnarray}
P(x,\bar{t}) &=&\Po(x) e^{-(\lambda+r)\bar{t}} \uu_{x\leqslant z_0}\nonumber \\
&+&\lambda \int^{\bar{t}}_{0} \dd \bar{t}' e^{-(\lambda+r)(\bar{t}-\bar{t}')}\Bigg[\int^{z(\bar{t}')}_{-\infty} h(y-x)\Po(y)\dd y\nonumber \\
&+&\int^{\infty}_{z(\bar{t}')} h(y-x)P(y,\bar{t}')\dd y\Bigg],
\label{AVanillaP}
\end{eqnarray}
where $z(\bar{t})$ fulfills 
$P(z(\bar{t}),\bar{t}) =\Po(z(\bar{t}))$,
and obviously $z_0\equiv\lim_{\bar{t} \rightarrow 0}z(\bar{t})$, 
$z_0\leqslant \xs$. The origin of this equation can be explained as follows. The holder of an American option must continuously decide the price of the underlying that triggers the early exercise, this is just the role that plays $z(\bar{t})$. Once this function is known, the pricing strategy must consider three excluding possibilities: 
\begin{enumerate}
\item
there is no jump prior to maturity, and the price (up to the discounting exponential factor) is $\Po(x)$ only if the present price is below $z_0$; 
\item
there is at least a jump that brings the asset price {\it below\/} the threshold, the option is exercised, and therefore the option price depends again on the pay-off function; 
\item
the stock price is still {\it above\/} the threshold, and the option remains alive. 
\end{enumerate}

\begin{figure}
{\hfil 
\includegraphics[width=1.0\columnwidth,keepaspectratio=true]{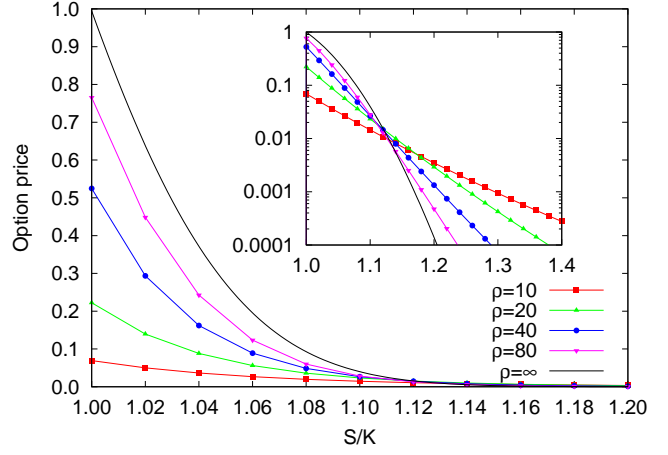}}
\caption{Price for an American binary put option. We have kept the same parameters as in fig.~\ref{Fig1}. The inset shows a different decay behaviour for every level of market activity.}
\label{Fig5}
\end{figure}

The core of the problem lies in the fact that in general one must find $P(x,\bar{t})$ and $z(\bar{t})$ {\it simultaneously\/}~\cite{McK65}.
However, there are exceptions to this rule, as in the case of binary puts, because for them $z(\bar{t})=\xs$.  
As a result,
the Laplace transform of $P(x,\bar{t})$ can be computed
\begin{eqnarray*}
\hat{P}(x,s) = 
\frac{\lambda}{\lambda+r+s}\left\{ \frac{1}{s} + \int^{\infty}_{\xs}\dd y h(y-x)\left[ \hat{P}(x,s)-\frac{1}{s}\right]\right\}, 
\end{eqnarray*}
as well as the equivalent differential equation when $h(x)$ is again described by eq.~(\ref{h}):
\begin{eqnarray*}
\partial^2_{xx}\hat{P}(x,s)+(\gamma-\rho) \partial_{x}\hat{P}(x,s)-\frac{r+s}{\lambda+r+s} \gamma \rho \hat{P}(x,s)=0. 
\end{eqnarray*}
Here we must solely investigate the solution for $x>\xs$, since $P(x,\bar{t})=1$ for $x \leqslant \xs$. The upper boundary condition $\lim_{x \rightarrow +\infty} P(x,\bar{t})=0$ leads to:
\begin{eqnarray*}
\hat{P}(x,s)=A(s)e^{\beta_- x}, 
\end{eqnarray*}
and we must use again the integral equation to get $A(s)$, since the price is discontinuous for $x=\xs$: 
\begin{eqnarray*}
A(s)=\frac{\gamma+\beta_-}{\gamma}\frac{e^{-\beta_- \xs}}{s}. 
\end{eqnarray*}
From this expression we can directly obtain the value of perpetual ($T\rightarrow \infty$) American puts because:
\begin{eqnarray*}
P(x,\bar{t}\rightarrow \infty)=\lim_{s\rightarrow 0} s \hat{P}(x,s)=\frac{\rho-1}{\gamma}e^{-(\gamma-\rho+1)(x-\xs)}, 
\end{eqnarray*}
a result already published in~\cite{M08}. In spite of the apparent simplicity of $A(s)$ the expression of $P(x,\bar{t})$ is very intricate: 
\begin{eqnarray*}
P(x,\bar{t})=\sqrt{\frac{2 \rho \lambda\bar{t}}{\gamma}}\int^{\infty}_0 \dd \xi I_1\left(\sqrt{2\gamma \rho \lambda\bar{t}} \xi\right) e^{-(\lambda+r)\bar{t}}\\ \times 
\Big[  \LL^{\gamma+1}_{\rho-1}(\xs-x;\xi)+\LL^{\rho-1}_{\gamma+1}(\xs-x;\xi)-\LL^0_{\gamma+\rho}(\xs-x;\xi)\Big],\\
\LL^a_b(x,\xi)=b e^{(a-\rho)x}\MM^a_b(x;\xi).
\end{eqnarray*}
In the continuous trading limit we can simplify the previous expression and recover once again the Wiener result~\cite{RR91}, but in the rest of the cases the wisest procedure is perhaps to invert numerically the Laplace solution, as we have done in the confection of fig.~\ref{Fig5}.~\footnote{In fact, all the figures in this paper were made by means of numerical inversion of either Laplace or Fourier expressions.}

We will finish the article with some discussion concerning the problem of a more general pay-off function for which $z(\bar{t})$ is not a constant, as for American vanilla put options, $\Po(x) =(K-e^{x}) \uu_{x\leqslant \xs}$. It is notorious that no closed expression is known in this case, even when the evolution of the return is driven by a Wiener process. Therefore we will present partial results only. For instance, we can compute the value of $z_0$, because it must solve
\begin{equation*}
\Po(z_0)=\frac{\lambda}{\lambda+r} \int^{+\infty}_{-\infty} h(y-z_0) \Po(y) \dd y.
\end{equation*}
When $h(x)$ follows eq.~(\ref{h}) and $\Po(x) =K-e^{x}$ we will have
\begin{eqnarray*}
Z_0=e^{z_0}=K  \left[ \frac{(\gamma+\rho)(\gamma-\rho+1)}{\gamma(\gamma+1)}\right]^{\frac{1}{\rho}}\leqslant K.
\end{eqnarray*}
In particular if $\rho \rightarrow 1$ we have $Z_0\rightarrow K$. And we recover the same result for the continuous trading limit $\lambda \rightarrow \infty$.

We may also compute the renewal equation for perpetual put prices~\cite{BL02b}, when $x>z^*$, with a general pay-off function, because it follows from formula~(\ref{AVanillaP}) that:
\begin{eqnarray*}
\Ps(x) =\frac{\lambda}{\lambda+r}\Bigg[\int^{z^*}_{-\infty} h(y-x)\Po(y)\dd y\\
+\int^{\infty}_{z^*} h(y-x)\Ps(y)\dd y \Bigg],
\end{eqnarray*}
where $z^*\equiv\lim_{\bar{t}\rightarrow \infty} z(\bar{t})$, and $\Ps(x)\equiv\lim_{\bar{t}\rightarrow \infty} P(x,\bar{t})$. The ordinary differential equation for $\Ps(x)$, when $h(x)$ follows~(\ref{h}), is
\begin{eqnarray*}
\Ps''(x)+(\gamma-\rho) \Ps'(x)-(\gamma-\rho+1)\Ps(x)=0, 
\end{eqnarray*}
whose solution reads:
\begin{eqnarray*}
\Ps(x)=(\rho-1)\left[\int^{0}_{-\infty} \Po(y+z^*) e^{\gamma y} \dd y\right]e^{(\gamma-\rho+1)(z^*-x)},
\end{eqnarray*}
where we have used the integral equation as well. The value of $z^*$ is obtained by demanding 
$\Ps(z^*) =\Po(z^*)$. In particular when we price vanilla puts
we have~\cite{M08}:
\begin{eqnarray*}
\Ps(x)=\frac{\rho-1}{\gamma}\left[K-\frac{\gamma}{\gamma+1}e^{z^*}\right]e^{(\gamma-\rho+1)(z^*-x)}, \\
Z^*=
e^{z^*}= K \frac{(\gamma+1)(\gamma-\rho+1)}{\gamma(\gamma-\rho+2)}.
\end{eqnarray*}
In the $\rho \rightarrow \infty$ limit we will obtain once again the Wiener results~\cite{IJK90}, 
\begin{eqnarray*}
Z^*=\frac{2r}{2r+\sigma^2}K,\qquad \Ps(x)=\frac{\sigma^2}{2r+\sigma^2}K e^{2 r(z^*-x)/\sigma^2}.
\end{eqnarray*}

\section{Conclusions}
\label{sect_end}
Along this paper we have introduced a very natural way of addressing the question of pricing financial derivatives within the framework of simple CTRW market models: the use of renewal equations. This approach clarifies the procedure to be followed if one wants to extend CTRW results coming from some other field of science to quantitative finance.

We have illustrated the potentials of this methodology by presenting its outcomes for standard contract specifications: European binary calls, European vanilla calls, American binary puts and (perpetual) American vanilla puts. For this purpose we have chosen a particular but exemplifying jump density: an asymmetric two-sided exponential function. 

The different results we have thus obtained exhibit those properties we expect to find in this kind of market models, like the observed discontinuities in the pricing expressions that capture the no-change likelihood which is inherent to the process, or the volatility smile. Nonetheless, the model is amenable enough to recover the celebrated Merton-Black-Scholes formulae for the Wiener process under the continuous trading limit. These capabilities might allow us to obtain new insights in the open problem of pricing American vanilla puts in the future. 

We have also analysed the case in which the pay-off function of an European derivative is Fourier transformable. In such a case a formal general pricing expression can be found, and therefore we may compare the outcomes for well-different jump densities. The conclusion is that $h(x)$ may play a significant role even when we approach to actual market conditions, in particular if we consider a density that shows slow convergence to a Gaussian.

Finally, in a forthcoming work we plan to compare systematically the key properties that option prices inherit from our market model with empirical evidences, coming from actual tick-by-tick data series, in order to decide when this approach becomes relevant o even compulsory. 

\begin{acknowledgement}
\noindent
The author acknowledges partial support from the MEC under contract No. FIS2006-05204-E. He is also grateful to Jaume Masoliver for his comments on the manuscript. 
\end{acknowledgement}

\appendix
\renewcommand{\theequation}{\Alph{section}.\arabic{equation}}
\renewcommand{\thesection}{Appendix \Alph{section}:}
\setcounter{equation}{0}
\setcounter{section}{0}
\section{Martingales and risk-neutral measures}

Let us set $t_0=0$ hereafter for the sake of algebraic simplicity, and recall that $\SI \equiv S(t_0)>0$. The stochastic differential equation for $S(t)$ 
if the waiting times are exponentially distributed is~\cite{M76,N93}
\begin{equation}
\frac{\dd S(t)}{S(t-)}=\left(e^{\Delta X(t)}-1\right) \dd N(t),
\label{SDE_S}
\end{equation}
where $N(t)$ is a right-continuous Poisson counting process, the number of jumps up to time $t$, whose increments are independent of $\Delta X(t)$. The mean value of $N(t)$ computed at time $t'$, under the {\it physical measure\/} $\Pm$, is 
\begin{equation}
\EE^{\Pm}[N(t)|\FF(t')]=N(t')+\lambda (t-t')\quad (0\leqslant t'\leqslant t),
\label{meanN}
\end{equation}
because $N(t')$ is a $\FF(t')$ measurable magnitude. The inverse of the mean jump time, $\lambda$, is commonly referred in mathematical textbooks as the {\it intensity\/} of the Poisson process. This implies that $N(t)$ is not a martingale with respect to the filtration $\FF(t)$ and the measure $\Pm$ because it is not fulfilling the conditional expectation property, which states that for any martingale $M(t)$ it is true that
\begin{equation}
\EE^{\Pm}[M(t)|\FF(t')]=M(t') \quad  (0\leqslant t'\leqslant t),
\label{martingaleG}
\end{equation}
and in particular that
\begin{equation*}
\EE^{\Pm}[dM(t)|\FF(t-)]=0.
\end{equation*}
From~(\ref{SDE_S}) and the independence of jump sizes and sojourns, one can see that
\begin{equation*}
\EE^{\Pm}[\dd S(t)|\FF(t-)]=S(t-)\left(\EE^{\Pm}[e^{\Delta X(t)}|\FF(t-)]-1\right) \lambda \dd t,
\end{equation*}
and then process $S(t)$ is not a martingale under the physical measure $\Pm$ with the filtration $\FF(t)$ unless we have that $\EE^{\Pm}[e^{\Delta X(t)}|\FF(t-)]=\EE^{\Pm}[e^{\Delta X(t)}]=\tilde{h}(-i) = 1$. In such a case we might even replace the Poisson process $N(t)$ with a different right-continuous counting process with bounded activity, and the martingale condition~(\ref{martingaleG}) would still hold for $S(t)$.

Therefore, in the most general case, when $\tilde{h}(-i) \neq 1$, $N(t)$ and $S(t)$ are not martingales but {\it semimartingales\/}: the sum of a martingale and a finite variation process. In order to prove  this statement let us introduce the compensated Poisson process $\bar{N}(t)\equiv N(t)-\lambda t$, and use expression~(\ref{meanN}) to show that it fulfills~(\ref{martingaleG}). Since we have $\EE^{\Pm}[|\bar{N}(t)|]<\infty$, one concludes that $\bar{N}(t)$ is a martingale, and as a direct consequence, that $N(t)$ is a semimartingale. 
$S(t)$ is a semimatingale as well: let us rewrite~(\ref{SDE_S}) in the following form:
\begin{eqnarray*}
\dd S(t)&=&S(t-)\left(e^{\Delta X(t)}-1\right)\lambda \dd t \\
&+& S(t-)\left(e^{\Delta X(t)}-1\right) \dd \bar{N}(t).
\end{eqnarray*}
The first term is a finite variation process if we have $\EE^{\Pm}[e^{\Delta X(t)}]=\tilde{h}(-i)  <\infty$, and the second one is a martingale because is the product of a martingale and a process that is adapted to it. 

For pure discontinuous semimartingales, the associated stochastic integral for~(\ref{SDE_S}), in the It\^o sense, reads:
\begin{eqnarray}
S(t)&=&\SI+\int^t_0 S(t'-)\left(e^{\Delta X(t')}-1\right) \dd N(t')\nonumber \\&=&
\SI+\sum^{N(t)}_{n=1} S(t_{n-1})\left(e^{\Delta X(t_n)}-1\right),
\label{sumS}
\end{eqnarray}
where it is understood that the summatory term vanishes whenever $N(t)=0$ and the solution is almost surely unique |both considerations apply to forthcoming expressions. This result is a direct consequence of the Dol\'eans-Dade exponential formula for semimartingales~\cite{CT04,K97}. Moreover, one can rewrite~(\ref{sumS}) in the more revealing form
\begin{eqnarray*}
S(t)=\exp\left\{x_0+\sum^{N(t)}_{n=1} \Delta X(t_n)\right\}=e^{X(t)},
\end{eqnarray*}
because $\dd X(t)=\Delta X(t) \dd N(t)=\Delta X(t) \lambda \dd t+ \Delta X(t) \dd \bar{N}(t)$, and then, in the It\^o sense
\begin{eqnarray*}
X(t)=x_0+\int^t_0 \Delta X(t') \dd N(t')= x_0+\sum^{N(t)}_{n=1} \Delta X(t_n),
\label{sumX}
\end{eqnarray*}
where we recall that $x_0\equiv\ln \SI$. The same result can be obtained by using the It\^{o}'s lemma for Poisson processes~\cite{K97,M71}:
\begin{eqnarray}
\dd F(x,t)&=&\partial_t F(x,t) \dd t+ \left[F(x+\Delta X(t),t)-F(x,t)\right] \dd N(t)\nonumber \\
&=&\left[\partial_t F(x,t) +\lambda  \mathcal{O}_x F(x,t)\right]\dd t+ \mathcal{O}_x F(x,t) \dd \bar{N}(t)\nonumber \\ 
&=&\partial_t F(x,t) \dd t+ \mathcal{O}_x F(x,t) \dd N(t),
\label{ItoLemma}
\end{eqnarray}
 where $\mathcal{O}_x$ is the following differential operator 
\begin{equation*}
\mathcal{O}_x\equiv \exp\{\Delta X(t) \partial_x\}-1 =\sum_{m=1}^{\infty} \frac{[\Delta X(t)]^m}{m!} \partial^{m}_{x},
\end{equation*}
and $x=X(t-)$.

Martingales are interesting magnitudes in finance because future (expected) values coincide with present values, which means that all the available information about the future price of any security is already included in the actual quoted price: the efficient market hypothesis. If one takes into account the natural depreciation that introduces the existence of a risk-free interest rate, this means that for any security $Y(t)$
\begin{equation}
\EE^{\Pm}[Y(t)e^{-r t}|\FF(t')]=Y(t') e^{-r t'}\quad (0 \leqslant t'\leqslant t),
\label{martingaleY}
\end{equation}
should hold. But this is not true in general because:
\begin{eqnarray*}
\EE^{\Pm}[S(t)e^{-r (t-t')}|\FF(t')]&=&S(t') e^{-(r-\lambda[\tilde{h}(-i)-1])(t-t')}\\
&\neq& S(t'), 
\end{eqnarray*}
if $r\neq \lambda[\tilde{h}(-i)-1]$.~\footnote{This means that if $\tilde{h}(-i)=1$, then $r=0$. Let us recall that is such a case $S(t)$ will be a martingale under $\Pm$ even when sojourns are not exponentially distributed.} 
The reason lies in the fact that securities are risky objects in the most of the cases, this is the main practical motivation to negotiate with them: the expectation of overcoming the evolution of risk-free assets for which the martingale property~(\ref{martingaleY}) holds. But this risk premium must be priced and investors may have different perception of and tolerance to risk. This lack of consensus is very relevant when pricing a derivative, since at the end its value is riskless fixed by the pay-off function at exercise. Then, it seems rational to demand derivative prices to be risk neutral.    

From a mathematical point of view, this demand can be satisfied if we change the measure of the probability space from the {\it physical\/} measure $\Pm$ to the {\it risk-neutral\/} measure $\Qm$, so that
\begin{equation}
\EE^{\Qm}[Y(t)e^{-r t}|\FF(t')]=Y(t')e^{-r t'}\quad (0 \leqslant t'\leqslant t),
\label{martingaleQ}
\end{equation}
holds now for any security, but in particular for the underlying asset $S(t)$ and any derivative $C(x,t)$. 
Every measure that fulfills~(\ref{martingaleQ}) is a valid candidate to be a risk-neutral measure. However, the usual practice is to choose a new measure that does not change the nature of the involved stochastic processes: the compensated Poisson process in the present case. The problem of defining the risk-neutral measure is equivalent to the search for the so-called state price process $\xi(t)$ that fulfills
\begin{equation*}
\EE^{\Pm}[\xi(t)Y(t)e^{-r t}|\FF(t')]=\EE^{\Qm}[Y(t)e^{-r t}|\FF(t')].
\end{equation*}
$\xi(t)$ is nothing but the Radon-Nikodym derivative of the measure $\Qm$ with respect to the measure $\Pm$:
\begin{equation*}
\xi(t)=\EE^{\Pm}\left[\left. \frac{\dd \Qm}{\dd \Pm}\right|\FF(t)\right],
\end{equation*}
and it is a martingale for both the physical and the risk-neutral measure. Then, by virtue of the representation theorem for pure jump processes,
\begin{equation*}
\frac{\dd \xi(t)}{\xi(t-)}=\eta(t) \dd \bar{N}(t),
\end{equation*}
for a given (predictable) process $\eta(t)$. The process $\eta(t)$ must be chosen therefore in such a way that $\xi(t) S(t) e^{-rt}$ is a martingale under the {\it physical\/} measure:
\begin{equation}
\dd\EE^{\Qm}\left[\left. S(t) e^{-rt}\right|\FF(t-)\right]=\dd \EE^{\Pm}\left[\left. \xi(t) S(t) e^{-rt}\right|\FF(t-)\right]=0.
\label{martingaleA}
\end{equation}
It\^o's calculus prescribes that,
\begin{eqnarray}
&&\frac{\dd \left[\xi(t) S(t) e^{-r t} \right]}{\xi(t-) S(t-) e^{-r t}}= \left[-r+\lambda(\eta(t)+1)\left(e^{\Delta X(t)}-1\right)\right]\dd t \nonumber \\
&&+\left[\eta(t)+(\eta(t)+1)\left(e^{\Delta X(t)}-1\right)\right] \dd \bar{N}(t),
\label{ItoS}
\end{eqnarray}
and therefore it suffices that 
\begin{equation}
\eta(t)\equiv -1 +\frac{r}{\lambda\left(e^{\Delta X(t)}-1\right)}, 
\label{exact}
\end{equation}
in order to fulfill martingale property~(\ref{martingaleA}). Observe that condition~(\ref{exact}) sets the value of $\eta(t)$ with no ambiguity, but requires $\Delta X(t)$ to be a predictable process, which is not our case in general. Therefore $\eta(t)$ cannot be assessed in that way. This fact implies that we will be not able to wipe eventually all the risk off the option price, no hedging portfolio can be defined, and the market becomes incomplete.

Note however that condition~(\ref{exact}) is not a {\it necessary\/} condition since we can define $\Qm$ through any $\eta(t)$ that fulfills  
\begin{equation*}
\EE^{\Pm}\left[\left.-r+\lambda(\eta(t)+1)\left(e^{\Delta X(t)}-1\right)\right|\FF(t-)\right]=0,
\end{equation*}
because second term in the RHS of eq.~(\ref{ItoS}) does not contribute in the computation of condition~(\ref{martingaleA}) |recall that jumps and sojourns were independent. Therefore we find that if $\tilde{h}(-i)\neq 1$
\begin{equation}
\eta\equiv \EE^{\Pm}\left[\eta(t)|\FF(t-)\right]=-1+\frac{r}{\lambda\left[\tilde{h}(-i)-1\right]},
\label{martingaleB}
\end{equation}
must hold, which only determines the mean value of $\eta(t)$. Therefore, any predictable process in the set of processes sharing condition~(\ref{martingaleB}) is a valid solution to the problem. In particular we can freely choose $\eta(t)=\eta$, as we have done in the rest of the paper.  Let us see next the implications that this choice conveys to option pricing. Like in the case of the stock price, $\xi(t) \CC(x,t) e^{-rt}$ should be a martingale, and the It\^o's lemma in~(\ref{ItoLemma}) leads to,
\begin{eqnarray}
\frac{\dd \left[\xi(t) \CC(x,t) e^{-r t} \right]}{\xi(t-) e^{-r t}}=\nonumber \\
\left[-r \CC(x,t)+\partial_t \CC(x,t)+\lambda(\eta+1) \mathcal{O}_x \CC(x,t)\right]\dd t \nonumber \\
+\left[\eta \CC+(\eta+1)\mathcal{O}_x \CC(x,t) \right] \dd \bar{N}(t).
\label{ItoC}
\end{eqnarray}
Therefore condition $\dd \EE^{\Pm}\left[\left. \xi(t) \CC(x,t) e^{-r t}\right|\FF(t-)\right]=0$ leads to a partial differential equation of infinite order:
\begin{eqnarray}
 \partial_t \CC(x,t)= r \CC(x,t)\nonumber \\
-\frac{r}{\tilde{h}(-i)-1}\sum^{\infty}_{m=1}\frac{(-i)^m}{m!}\partial^{m}_x \CC(x,t)\left.\partial^{m}_\omega \tilde{h}(\omega)\right|_{\omega=0},
\label{PDE} 
\end{eqnarray}
where we have used eq.~(\ref{martingaleB}). Note that the physical parameter $\lambda$ does not appear in this equation. Finally we can compare (\ref{ItoC}) with $\dd \left[\CC(x,t) e^{-r t} \right]$,
\begin{eqnarray*}
\frac{\dd \left[\CC(x,t) e^{-r t} \right]}{e^{-r t}}&=& \left[-r \CC(x,t)+\partial_t \CC(x,t)+\lambda \mathcal{O}_x \CC(x,t)\right]\dd t \nonumber \\
&+&\mathcal{O}_x \CC(x,t) \dd \bar{N}(t),
\end{eqnarray*}
and conclude that if we re-define the value of $\lambda$ to fulfill eq.~(\ref{lambda}), {\it i.e.\/} if we demand $\eta=0$, the equations obtained by using the physical measure coincide with those derived from a risk-neutral measure. It is not difficult to derive (\ref{PDE}) from (\ref{ERenewalBL}).


\begin{thebibliography}{000}
\bibitem{MW65} \Journal{E.W. Montroll and G.H. Weiss}{J. Math. Phys}{6}{167}{1965}
\bibitem{CM65} D.R. Cox and H.D. Miller, {\it The Theory of Stochastic Processes\/} (Wiley, New York, 1965)
\bibitem{GUY24} \Journal{G.U. Yule}{Philos. Trans. R. Soc. B-Biol. Sci.}{213}{21}{1925}
\bibitem{WF50} W. Feller, {\it An Introduction to Probability Theory and its Applications\/} (Wiley, New York, 1950)
\bibitem{W94} G.H. Weiss, {\it Aspects and Applications of the Random Walk\/} (North-Holland, Amsterdam, 1994)
\bibitem{MMW03} \Journal{J. Masoliver, M. Montero, and G.H. Weiss}{Phys. Rev. E}{67}{021112}{2003}
\bibitem{MMPW06} \Journal{J. Masoliver, M. Montero, J. Perell\'o, and G.H. Weiss}{J. Econ. Behav. Organ.}{61}{577}{2006}
\bibitem{ES06} \Journal{E. Scalas}{Physica A}{362}{225}{2006}
\bibitem{C62} D.R. Cox, {\it Renewal Theory\/} (Methuen, London, 1962)
\bibitem{W98}  P. Wilmott, {\it Derivatives}, (Wiley, Chichester, 1998) 
\bibitem{CR76}  \Journal{J.C. Cox and S.A. Ross}{J. Financ. Econ.}{3}{145}{1976}
\bibitem{M76}  \Journal{R.C. Merton}{J. Financ. Econ.}{3}{125}{1976}
\bibitem{CT04} R. Cont and P. Tankov, {\it Financial Modelling With Jump Processes\/} (CRC, London, 2004)
\bibitem{AS04} \Journal{Y. A\"{\i}t-Sahalia}{J. Financ. Econ.}{74}{487}{2004}
\bibitem{ASSW06} \Journal{S. Albeverio, M. Schmitz, V. Steblovskaya, and K. Wallbaum}{Stoch. Anal. Appl.}{24}{241}{2006}
\bibitem{BL02} S.I. Boyarchenko and S.Z. Levendorski\v{\i}, {\it Non-Gaussian Merton-Black-Scholes theory\/} (World Scientific, Singapore, 2002)
\bibitem{R07} \Journal{N. Ratanov}{Quant. Financ.}{7}{575}{2007}
\bibitem{W91} D. Williams, {\it Probability with martingales\/} (Cambridge University Press, Cambridge, 1991) 
\bibitem{SGM00} \Journal{E. Scalas, R. Gorenflo, and F. Mainardi}{Physica A}{284}{376}{2000}
\bibitem{CC07} \Journal{\'{A}. Cartea and D. del-Castillo-Negrete}{Physica A}{374}{749}{2007}
\bibitem{GPAMS99} \Journal{P. Gopikrishnan, V. Plerou, L.A.N. Amaral, M. Meyer, and H.E. Stanley}{Phys. Rev. E}{60}{5305}{1999}; \Journal{V. Plerou, P. Gopikrishnan, L.A.N. Amaral, M. Meyer, and H.E. Stanley}{ Phys. Rev. E}{60}{6519}{1999}
\bibitem{RE05} \Journal{J.R. Russell and R.F. Engle}{J. Bus. Econ. Stat.}{23}{166}{2005}
\bibitem{SY07} \Journal{A.C. Silva and V.M. Yakovenko}{Physica A}{382}{278}{2007}
\bibitem{BS73} \Journal{F. Black and M. Scholes}{J. Pol. Econ.}{81}{637}{1973}
\bibitem{MS94} \Journal{R.N. Mantegna and H.E. Stanley}{Phys. Rev. Lett.}{73}{2946}{1994}
\bibitem{K95} \Journal{I. Koponen}{Phys. Rev. E}{52}{1197}{1995}
\bibitem{CGMY02} \Journal{P. Carr, H. Geman, D. Madan, and M. Yor}{J. Bus.}{75}{305}{2002}
\bibitem{BL00} \Journal{S.I. Boyarchenko and S.Z. Levendorski\v{\i}}{Int. J. Theor. Appl. Financ.}{3}{549}{2000} 
\bibitem{BL02b} \Journal{S.I. Boyarchenko and S.Z. Levendorski\v{\i}}{SIAM J. Control Optim.}{40}{1663}{2002} 
\bibitem{ML07} \Journal{M.C. Mariani and Y. Liu}{Physica A}{377}{590}{2007}
\bibitem{ER98} \Journal{R.F. Engle and J.R. Russell}{Econometrica}{66}{1127}{1998}
\bibitem{McK65} \Journal{H.P. McKean}{Ind. Manag. Rev.}{6}{32}{1965}
\bibitem{M08} \Journal{M. Montero}{Physica A}{387}{3936}{2008}
\bibitem{RR91} \Journal{M. Rubinstein and E. Reiner}{Risk}{4}{75}{1991}
\bibitem{IJK90} \Journal{I.J. Kim}{Rev. Financ. Stud.}{3}{547}{1990}
\bibitem{N93}  \Journal{V. Naik}{J. Financ.}{48}{1969}{1993}
\bibitem{K97} O. Kallenberg, {\it Foundations of Modern Probability\/} (Springer-Verlag, New York, 1997)  
\bibitem{M71}  \Journal{R.C. Merton}{J. Econ. Theory}{3}{373}{1971}
\end{thebibliography}
\end{document}